\documentstyle[preprint,aps]{revtex}
\newcommand{\be}{\begin{equation}}
\newcommand{\ee}{\end{equation}}
\tighten
\draft
\title{Quantum Breaking of Elastic String}
\author{L. S. Levitov$^{a,b}$, A. V. Shytov$^b$, and A. Yu. Yakovets$^b$}
\address{(a) Massachusetts Institute of Technology,
77, Massachusetts Avenue, Cambridge MA02139,\\
(b) L. D. Landau Institute for Theoretical Physics,
2, Kosygin str., Moscow 117334, Russia}

\begin{document}

\maketitle

\begin{abstract}
Breaking of an atomic chain under stress is a collective
many-particle tunneling phenomenon. We study classical dynamics
in imaginary time by using conformal mapping technique, and
derive an analytic formula for the probability of breaking. The
result covers a broad temperature interval and interpolates
between two regimes: tunneling and thermal activation. Also, we
consider the breaking induced by an ultrasonic wave propagating
in the chain, and propose to observe it in an STM experiment.
  \end{abstract}
\narrowtext
  \noindent{\it Introduction.}\hskip4mm
Among various examples of quantum decay in condensed
matter~\cite{KaganLeggett} of special interest are those where
the tunneling system is strongly coupled to a macroscopic system.
Several such systems were studied,
including metastable current states of Josephson
junctions~\cite{criticalcurrent}, transport in tunnel
junctions~\cite{tunneljunction}, quantum diffusion
and properties of two-level
systems coupled to thermal bath~\cite{twolevel}. Our
purpose here is to discuss yet another system where decay is a
many-particle effect: breaking of elastic string stretched by an
external force. Our central example  will  be  a  polymer  chain
teared  under  stress,  however the formalism we present is more
general. For instance,  it  describes  critical  velocity  of  a
superfluid   flow   in   a  long  channel  and  breakdown  of  a
supercurrent carried by a thin  wire  or  by  many  small
capacitance Josephson junctions connected in series.

For the string, the leading contribution to the tunneling
exponent is due to a collective motion during
breaking~\cite{Dyakonov}. According to the Caldeira-Leggett
theory~\cite{CaldeiraLeggett}, since the string corresponds to
their ``ohmic'' case and is described by a harmonic Hamiltonian,
the internal degrees of freedom can be integrated out to produce
an effective action for which a classical trajectory in
imaginary time has to be determined. Although in principle there
is no difficulty in carrying out this program, we chose a
different method. The ``collective motion'' is just a
propagation of acoustic waves, and it can be treated exactly: in
imaginary time the wave equation becomes two dimensional Laplace's equation
that can be solved by conformal transformations. The conformal mapping method
gives a clear intuitive picture of the entire system motion in
imaginary time and allows to treat in parallel several
related problems: tunneling at zero temperature and at finite
temperature, and also tunneling caused by an alternating
tension. The latter problem is of a particular practical
interest because it corresponds to the breaking induced by an
ultrasonic wave propagating in a polymer.

We consider the limit of small tension, when the
tunneling results from coherent motion of many atoms. A simple
argument that allows to estimate the number of moving atoms and
the exponent in the breaking probability was given by
Dyakonov~\cite{Dyakonov}. During the time of tunneling $t$,
acoustic waves propagate at the distance $ct$, where $c$ is the
sound velocity. At small tension, the time $t$ is big, and thus
the mass involved in the tunneling, $m_*=\rho ct$, is much
bigger than the single atom mass. (Here $\rho$ is the
density of the chain.) One can then describe dynamics
approximately as tunneling of the mass $m_*$ under a
triangular barrier $U(x)=E_0-fx$, where $E_0$ is the energy of
the broken bond and $f$ is the tension. Then it is straightforward
to estimate the action $S\simeq f^{-1}m^{1/2}_* E^{3/2}_0$ and
the time $t\simeq f^{-1}m^{1/2}_*E^{1/2}_0$. By solving the
latter equation together with the selfconsistency relation
$m_*=\rho ct$, one finds $t\simeq \rho c E_0/f^2$, $m_*\simeq
\rho^2 c^2 E_0/f^2$, which gives the probability of breaking
$W\approx\exp(-{\rm const} E^2_0\rho c/\hbar f^2)$. The argument
of Dyakonov shows that collective effects are essential, as
$m_*$ grows at small tension. In this form, however, it
only gives a rough estimate of $\ln W$,
leaving an unknown constant in the exponent to be
determined by a more precise calculation.

\noindent{\it Discussion of experiment.}\hskip4mm
Beyond a theoretical interest, the quantum problem of string
breaking is relevant for the physics of
polymers, since at low temperature this process
determines maximal stress $\sigma_{max}$ that a polymer material
can sustain. At high temperature the breaking is thermally
assisted: $\sigma_{max}(T) \sim \exp E_0/T$. At low temperature,
a saturation of the raise of $\sigma_{max}(T)$ was
reported\cite{saturation}, however, the relation with tunneling
has not been clarified. It would be of interest to check whether
the saturation indicates tunneling by using, e.g., the
conventional technique of isotopic substitution. Since $\ln
W$ scales as $1/\rho c$, there is an isotope effect in the quantum
regime: $\ln W\sim\sqrt{M}$, which disappears in the thermally
assisted regime. Also, one could look at breaking enhancement in
presence of an alternating stress caused, e.g., by a microwave
radiation. In this case, the characteristic feature to observe
is the threshold frequency $\Omega_0$ where the breaking
enhancement described by our Eq.(\ref{HighFrequency}) begins.
The threshold is set by inverse tunneling time $t^{-1}\simeq
f^2/\rho cE_0$, and thus it is a function of the tension $f$ and
can be much lower than Debye's frequency $\omega_D$, the characteristic
threshold for the frequency effect in the thermally assisted
regime.

Also, in the light of recent progress in manipulating single
atoms by a scanning tunneling microscope~\cite{STM}, it is of
interest to discuss an STM experiment with a single polymer
molecule suspended between the surface and the STM tip acting as
a probe. In this setup one can study the molecule breaking at
small and at large temperature, where our calculation predicts
different regimes. In addition, there is a possibility to
combine constant force with an AC tension field caused, e.g., by
an AC signal on the tip, and to look at resonance effects
related with exciting a standing ultrasonic wave in the
molecule. Due to a large acoustic mismatch with the surface,
reflection of the ultrasonic wave at the ends of the molecule is
almost perfect. Thus the molecule of length $L$ works like a
high quality phonon resonator with resonance frequencies
$\omega_n=\pi n c/L$ corresponding to sinusoidal standing waves.
For example, in a chain of 1000 atoms one estimates $\omega_1$
as $10^{-3}\omega_D\approx 5-10\ GHz$. A signal in this
frequency range presumably can be supplied by coupling the STM
tip to a microwave source.

Breaking then will occur at the nodes of the wave, where local
tension is maximal, and this will result in simple ratios of
broken halves: 1:1, 2:1, 3:2, etc. On the other hand, at high
frequency there is no particular spatial
structure of the breaking. Like in the bulk polymer, the experiment is
facilitated by anomalously low frequency threshold $\Omega_0\sim
t^{-1}$ above which the breaking is exponentially enhanced.

  \noindent{\it Result and method.}\hskip4mm
The string is described by the wave equation that
in imaginary time turns into Laplace's equation. We solve the
latter by a conformal mapping and derive a formula
for the probability of breaking
  \begin{equation}\label{mainresult}
W\ =\ {\Delta}\ \exp\left( -\frac{2E_0^2 \rho c}{\pi \hbar f^2} \right)
\ ,
\end{equation}
  where ${\Delta}\simeq \omega_De^{-E_0/\omega_D}$ is the probability of
a single atom tunneling at one atomic spacing. Let us note an
agreement with the qualitative estimate given before. We
generalize this result to a finite temperature and derive an
exact formula that covers a broad temperature range, including
thermal activation at high temperatures.

We shall use the following picture of tunneling. The energy
$U(x)$ of a bond falls off as its length $x$
exceeds several angstroms. Therefore, due to zero-point
fluctuations, the bond occasionally makes excursions in the state
where it is virtually broken, with the frequency
$\simeq {\Delta}$. If there were no tension, after every
such excursion the bond would return back to its equilibrium
length. However, in the presence of the force stretching the
chain, after the bond is virtually broken, the ends of the tear
can start moving apart in the classically forbidden region, until the
distance between them reaches $\simeq E_0/f$, where the barrier ends and the
bond is really broken. This second stage takes much longer than
${\Delta}^{-1}$ as it involves motion of many atoms at large
distance. Thus, we have two stages of tunneling: (1) fast
virtual disappearance of one bond; (2) slow collective motion
under applied force traversing classically forbidden region. After
we use this picture to do the calculation, we check
selfconsistency of the assumptions, and confirm validity of
the approach for the tension small on the atomic scale.

Clearly, since the first stage is essentially  one-particle  and
does  not  depend  on the tension, it only contributes a
prefactor
$\simeq {\Delta}^{1/2}$ of the exponentially small amplitude  due
to  the  second  stage.  The latter can therefore
be analyzed using  the  method   of   sudden
switching~\cite{suddenswitching}.    Let    us   consider   a
Hamiltonian
  $$
{\cal{H}} = \sum_{k=1...N } \left[\frac{\hat{p}_k^2}{2M} +
U(\hat q_k - \hat q_{k-1})\right]\ +\
U(\hat q_1)+
U(L - \hat q_N)
  $$
where $M$ is atomic mass, $U(x)$ is the interaction of
neighbors, $q_k$ are atoms' coordinates. The ends of the
chain are fixed at $x=0,L$, which determines interatomic spacing $a=L/(N+1)$
and the tension $f=U'(a)$. Our $f$ is small,
so $a$ is close to the equilibrium spacing $a_0$ for
which $U'(a_0)=0$, and we can use harmonic approximation
  \be
U(x) =-E_0+{1\over2}
m \omega_0^2 (x-a_0)^2
\ee
   for all bonds except the one that breaks. For the breaking
bond, we practically do not need exact profile of the potential
$U(x)$, as it is important only during a very short time
interval $\simeq {\Delta}^{-1}$ at the beginning of the tunneling.
Therefore, as we discussed, it contributes only to the
prefactor of Eq.(\ref{mainresult}). According to \cite{suddenswitching},
we write the rate of transition to the broken state as
  \be
\label{SuddenSwitch}
W={\Delta}\int\limits^\infty_{-\infty}\langle\exp(-{i\over\hbar}
\hat{\cal H}'t)\rangle dt,
\ee
  where the Hamiltonian $\hat{\cal H}'$ describes the chain with
one bond removed: $\hat{\cal H}'=\hat{\cal H}-U(\hat x_j-\hat
x_{j-1})-E_0$. (We subtract $E_0$ to assure energy conservation.)
The average in Eq.(\ref{SuddenSwitch}) is
taken over the initial state, i.e., the ground state of
$\hat{\cal H}$.

Following the usual treatment~\cite{pathintegral}, we consider the
tunneling
as a motion in imaginary time: first, we evaluate the average in
(\ref{SuddenSwitch}) at imaginary values of $t$, then
continue it from the upper half-plane down to the real
axis and do the integral over $t$ by the saddle point
method. To compute the average, we write
$\exp(-t \hat{\cal H}'/\hbar)$ averaged over the initial wave
function as the path integral
  \be\label{pathintegral}
\langle\exp(-{t\over\hbar}\hat{\cal H}')\rangle =e^{-E_0t/\hbar}\
\int {\cal D}[q]\exp\left(-{1\over\hbar}S_{it/2}[q]\right)
\ee
  Here $S_{i\tau}$ is the classical action of the chain with one bond
broken during the time interval $[-i\tau, i\tau]$:
  \be
S_{i\tau}[q]=
\int\left[ {\cal H}-\theta({\tau^2}-t^2)U(q_j-q_{j-1})\right] dt\ ,
\ee
  where $j$ is the number of the broken bond.
The integral (\ref{pathintegral}) is
taken over paths going from $t = - \infty$ to $t =
\infty$, i.e., over all states $\{q_k(t)\}$ of the chain. Since
at small tension the tunneling involves only long wavelength
distortions, we go to the continual limit:
  \be\label{c-limit}
S = \int\int \left[ \frac{\rho}{2} \dot u^2
+ \frac{\rho c^2}{2} u^2_x - f u_x \right] dx\,dt \ ,
\ee
   where $x(k)=a_0k$, and the atoms' coordinates $q_k$ are
written through the displacements relative to the equilibrium
positions: $q_k(t)= u(x,t)+a_0k$. Together with the boundary
condition corresponding to  the broken bond at $x=0$, this defines a
linear problem. We solve it and evaluate the functional integral
by the saddle point method.

Beginning from here we use the units $\hbar=\rho=c=1$, and
recover the usual units only when necessary. Classical equation
of motion obtained from Eq.(\ref{c-limit}) is
   \be
{\partial^2u\over\partial x^2}+
{\partial^2u\over\partial t^2}\ =\ 0\ .
\ee
  We assume that the breaking occurs far from the ends
and treat the chain as infinite. Then the Laplace's equation is
supplied with the boundary conditions: \hfil \break
  (i) $u(x,t)\to u_0(x,t)=fx$ at $|x|,|t|\gg {\tau}$, i.e.,
the chain is uniformly stretched away from the tear;\hfil \break
  (ii) $\partial u/\partial x = 0 $
at $x=\pm0$,
$-{\tau}<t <{\tau}$, i.e., the ends of the tear  at $x=0$
are free
during the tunneling.\hfil \break
By analogy with 2D electrostatics, the boundary value problem
can be solved by a proper conformal mapping. For that, let
us introduce complex variable $z=x+it$ and consider analytic
function
  \be\label{joukowsky}
w(z)=(z + \sqrt{z^2+{\tau}^2})/{\tau}\ .
\ee
   The cut $[-i{\tau},i{\tau}]$ in the $z-$plane is mapped on the unit
circle $|w|=1$, and thus we come to
Laplace's equation in the domain $|w|>1$ with the  boundary
conditions
   \be\label{w-cond}
{\rm (i)}\  u=u_0(w)\ {\rm at\ } |w|\gg1; \quad
{\rm (ii)}\ {\partial u\over\partial |w|}= 0 \  {\rm at\ } |w|=1
\ee
 It has a solution
\be
\label{instanton}
u(z) = {1\over2} f{\tau}\,{\rm Re\,}\left(w + \frac{1}{w}\right) =
f{\rm Re\,} \sqrt{z^2+{\tau}^2}\ .
\ee
  Due to conformal invariance, the action  can  be
evaluated  in  the  $w-$plane:  $  S_{i\tau}[u-u_0]  = -\pi f^2
{\tau}^2/2$. Then we continue $S_{i\tau}$ to the real axis,
substitute it  in
Eq.(\ref{SuddenSwitch}), do the integral
  \be\label{integral}
W\simeq \int d{\tau} \exp\left(-\frac{\pi}{2}f^2 {\tau}^2 -
2iE_0 {\tau}\right)
  \ee
and get Eq.(\ref{mainresult}) for the probability.

With the result (\ref{instanton}), we can check the
selfconsistency of the method. During the tunneling, the width
of the tear $u(x=+0,t)-u(x=-0,t)={2f\over\rho
c}\sqrt{{\tau^2}-t^2}$ is estimated as $E_0/f$, as the saddle
point evaluation of the integral (\ref{integral}) sets
${\tau}=2i\rho c E_0/\pi f^2$. If the tension $f$ is small in atomic
units, the tear is much wider than the interatomic spacing,
except for $t$ very close to $\pm {\tau}$, and thus indeed it can
be described by the free end boundary condition at the cut.

  \noindent{\it Finite temperature.}\hskip4mm
Temperature dependence of the breaking begins very early, at the
temperature set by the inverse tunneling time,
$T_c\simeq\hbar/\tau=\hbar f^2/\rho cE_0$. If the tension is weak,
$T_c$ can be much smaller than the usual temperature
$\hbar\omega_D$ of transition to the thermally assisted
tunneling. For such temperature the breaking
can be studied by a generalization of the zero
temperature calculation.

The functional integral treatment of tunneling at
finite temperature amounts to integrating over periodic
trajectories with the imaginary period $i\beta=i/T$~\cite{pathintegral}.
For the chain, we have to evaluate the path integral
Eq.(\ref{pathintegral}) taken over all functions $u(z)$ periodic
in the $z-$plane with the period $i\beta$: $u(z+i\beta)=u(z)$.
Repeating the steps that lead to Laplace's equation, we get a
boundary value problem for the function $u(z)$ whose
normal derivative vanishes near a periodic system of cuts, $z\in
[-i{\tau}+i\beta n,\ i{\tau}+i\beta n]$. (Condition at
$z=\infty$ remains unchanged.)

Corresponding conformal mapping can be constructed in two
successive steps. First, let us consider the function
$\eta(z)=\exp(2\pi Tz)$ that transforms the strip $0<{\rm
Im\,}z<\beta$ to the whole complex plane, and maps $z=\infty$ to
two points: $\eta=0$, $\eta=\infty$. The cut is mapped to the
arc of the unit circle that goes from $\alpha=\exp(2\pi
iT{\tau})$ to $\bar\alpha$ counterclockwise. The second function
we take is
  $$w(\eta)={\eta-\alpha\over \alpha\eta-1}\ . $$
In the $w-$plane the two images of $z=\infty$ are $w_1=\alpha$,
$w_2=\bar\alpha$, and the cut goes along the negative real half axis
$w<0$. Then the boundary value problem is
solved by
   $$u(w)={if\over2\pi T} \ln\left({
(\sqrt{w}-\sqrt{w_1}) (\sqrt{w}+\sqrt{w_2})\over
(\sqrt{w}-\sqrt{w_2}) (\sqrt{w}+\sqrt{w_1})}\right)$$
  with the standard branch of the square root. Computation of
the action $S_{i\tau}[u-u_0]$ is facilitated by comparing with analogous
problem of 2D magnetostatics: two opposite charge magnetic
monopoles situated at the points $w_{1,2}$ near a
superconducting slit going along the real axis, from $0$ to the
left. In this formulation, the action corresponds to the
interaction energy of the monopoles and their images.
Straightforward calculation gives
   $$S_{i\tau}[u-u_0]=\lambda\ln\cos^2(\pi T{\tau})\ ,\
\lambda={f^2\over2\pi T^2}\ .$$
   To obtain the probability $W$, we take $i{\tau}$ back to the real
axis, substitute $S_{\tau}$  in Eq.(\ref{SuddenSwitch}), and do the integral:
  \be\label{generalcase}
W\simeq\int{e^{2iE_0{\tau}}d{\tau} \over\cosh^{2\lambda}(\pi T{\tau})}=
{2^{2\lambda}\over2\pi T}
{\Gamma(\lambda_+) \Gamma(\lambda_-)\over
\Gamma(2\lambda)}\ ,
\ee
  $\lambda_\pm=\lambda\pm iE_0/\pi T$.
This expression covers a large
temperature range, up to $T\simeq\omega_D$.
By using Stirling formula
we rewrite the result:
 $$\ln W(f,T)=-{E_0\over\pi T}\left(2\tan^{-1}\xi-
{1\over\xi}\ln(\xi^2+1)\right)\ ,
$$
  where $\xi=2TE_0\rho c/\hbar f^2$.
Asymptotically, at high temperature, $\xi\gg1$, we get Arrhenius
law $ W\simeq\lambda e^{-E_0/T}$, and in the opposite limit, $\xi\ll1$,
Eq.(\ref{generalcase}) matches
Eq.(\ref{mainresult}).

\noindent{\it Breaking due to alternating tension.}\hskip4mm
Now, let us discuss how the breaking can be stimulated by a
sound wave. Local tension in the wave is varying along the
string, and so is the breaking probability.
To study the spatial dependence,
let us put a phase $\phi$ in the wave,
  \be\label{soundwave}
u_0(x,t) = \frac{f_0}{\rho c\Omega} \cos(\Omega(x/c - t)+\phi)\ ,
\ee
  where $f_0$ is the tension amplitude,
and then study breaking at $x=0$ at different values of $\phi$.
In the $z-$plane
  \be\label{z-wave}
u_0(z) = \frac{f_0}{\rho c\Omega} {\rm Re}\, e^{i(\kappa z-\phi)}\ ,
\ee
  where $\kappa=\Omega/c$. Let us assume zero temperature,  then
we  have  Laplace's  equation  for the displacement field $u(z)$
with the boundary conditions identical to that we had in the $T=0$
constant force problem.
By the conformal transformation Eq.(\ref{joukowsky}) we go
to the domain $|w|>1$ in the $w-$plane,
$z={\tau}(w-1/w)/2$, where we get the boundary value
problem (\ref{w-cond}) with $u_0(w)$ corresponding to Eq.(\ref{z-wave}).
Solution is readily obtained by writing  Laurent  series  for $u(w)$ and
$u_0(w)$,  solving separately for each term, and then collecting
the series:
  \be
u(w) = u_0(w)+
{2f_0\over\Omega}\sin\phi\,
{\rm Re}
\sum_{m=1}^{\infty}{i^{m+1}\over w^m}J_{m}(i\Omega {\tau})\ ,
\ee
   where $J_{m}(x)$ are Bessel functions. The
action is
  \be
 S_{i\tau}[u-u_0]= - 2\pi{f_0^2\over\Omega^2}\sin^2\phi
\sum_{m=1}^{\infty} m |J_m(i\Omega {\tau})|^2
\ee
This sum can be done in the following two cases.

I. (low frequency: $\Omega {\tau} \ll 1$) Here $J_n(x)=O(x^n)$, and
only the $m=1$ term  survives:
\be
S_{i\tau} = - \frac{\pi f_0^2 {\tau}^2}{2 \rho c} \sin^2\phi
\ee
  This expression is just what one has for constant tension with the
local value $f= f_0 \sin\phi$. So, in this limit the
breaking occurs predominantly at the nodes of the displacement, where
the tension is maximal.

II. (high frequency: $\Omega {\tau} \gg 1$) Here we use the
$x\gg m$ asymptotics, $ J_m(ix) \approx i^m e^x/\sqrt{2 \pi x}$,
and get
  \be
S_{i\tau} =  - \frac{ f_0^2\sin^2\phi}{\rho c \Omega}\tau e^{2 \Omega {\tau}}
  \ee
So, the probability of breaking in this limit is
  \be
\label{HighFrequency}
W \sim \exp\left[ -\frac{E_0}{\hbar \Omega}
\ln\left(\frac{E_0 \rho c \Omega}{f_0^2\sin^2\phi} \right) \right]
\ee
   Let us note a similarity between Eq.(\ref{HighFrequency}) and
the ionization rate of an atom in a low frequency electric field
$f\cos\Omega t$: $W \sim f^{2n}$, where $n ={\rm Ry}/\hbar \Omega$
is interpreted as the number of absorbed quanta and $f^2$ as the
single quantum absorption rate. This interpretation stands
meaningful for the string as well.

To summarize, the frequency dependence of the breaking
probability is as follows: at $\Omega \ll \Omega_0={\tau}^{-1}=
f^2_0/\rho cE_0$ it is given by (\ref{mainresult}), and
the breaking occurs at the nodes of the wave. At $\Omega \ge
\Omega_0$ the frequency dependence begins and the probability
sharply increases following
Eq.(\ref{HighFrequency}). The spatial dependence
practically vanishes in this limit, since many cycles are repeated during
the tunneling.

To conclude, we studied breaking of a polymer chain, and derived
an analytic expression for the breaking probability that
describes transition from quantum tunneling to Arrhenius law.
Also, we studied breaking induced by an ultrasonic wave, and
discussed opportunities for experiment with bulk polymers and
with an STM used as a probe.

\acknowledgements   We   are
grateful  to  M.I.Dyakonov  and  S.E.Korshunov  for  valuable  and
illuminating discussions. Research at the  Landau  Institute  is
supported  by  ISF  grant  \#M9M000.  Research  of L.L. is
supported by Alfred Sloan fellowship.

\end{document}